\documentclass[final,5p,times,twocolumn]{elsarticle}

\usepackage{graphicx}
\usepackage{amssymb}

\journal{Journal of Physics and Chemistry of Solids}

\begin{document}

\begin{frontmatter}

\title{Pseudogap phase of high-Tc compounds described within the LDA+DMFT+$\Sigma$ approach
}

\author[1]{I.A. Nekrasov\corref{cor1}}
\ead{nekrasov@iep.uran.ru}
\author[1]{E.Z. Kuchinskii}
\author[1]{M.V. Sadovskii}

\address[1]{Institute for Electrophysics, Russian Academy of Sciences, Ural Branch,
Ekaterinburg, Amundsen str. 106, 620016, Russia }

\cortext[cor1]{Corresponding author.}

\begin{abstract}

LDA+DMFT+$\Sigma_{\bf k}$ approach was applied to describe pseudogap phase of
several prototype high-Tc compounds e.g. hole doped Bi$_2$Sr$_2$CaCu$_2$O$_{8-\delta}$ (Bi2212)
and La$_{2-x}$Sr$_x$CuO$_4$ (LSCO) systems and electron doped Nd$_{2-x}$Ce$_x$CuO$_4$ (NCCO) and
Pr$_{2-x}$Ce$_x$CuO$_4$ (PCCO), demonstrating qualitative difference of the Fermi surfaces (FS) for these systems.
Namely for Bi2212 and LSCO the so called ``hot-spots'' (intersection of a bare FS and AFM Brillouin zone 
(BZ) boundary), where scattering on pseudogap fluctuations is most intensive were not observed.
Instead here we have Fermi arcs with smeared FS close to the BZ boundary. However for
NCCO and PCCO ``hot-spots'' are clearly visible. This qualitative difference is shown 
to have material specific origin. Good agreement with known ARPES data was demonstrated 
not only for FS maps but also for spectral function maps (quasiparticle bands including 
lifetime and interaction broadening).

\end{abstract}


\end{frontmatter}



\section{Introduction}

One of the most prominent phenomena in high-Tc cuprates physics is the so called pseudogap~\cite{psgap}.
Here we present an overview of our recent works Refs.~\cite{Bi2212,LSCO,NCCO_work,PCCO_work}
on LDA+DMFT+$\Sigma_{\bf k}$ computational scheme applcations.
This scheme is generalization of dynamical mean-field theory DMFT~\cite{georges96}
and LDA+DMFT~\cite{ldadmft} (LDA -- local density approximation) approach 
allowing to include non-local scale dependent effects.~\cite{cm05,FNT}
To include pseudogap fluctuations effects important for cuprate
physics we supplied (in additive manner) conventional DMFT
with an “external” k-dependent self-energy $\Sigma_{\bf k}$.
For the pseudogap state $\Sigma_{\bf k}$ describes the
interaction of correlated electrons with non-local (quasi) static
short-ranged collective Heisenberg-like AFM or SDW-like spin fluctuations~\cite{Sch,KS}.

Within LDA+DMFT+$\Sigma_{\bf k}$ approach several high-T$_c$ prototype compounds e.g.
hole doped Bi$_2$Sr$_2$CaCu$_2$O$_{8-\delta}$ (Bi2212)~\cite{Bi2212} and
La$_{2-x}$Sr$_{x}$CuO$_4$ (LSCO)~\cite{LSCO} as well as electron doped
Nd$_{2-x}$Ce$_x$CuO$_4$ (NCCO)~\cite{NCCO_work} and
Pr$_{2-x}$Ce$_x$CuO$_4$ (PCCO)~\cite{PCCO_work} were studied.
Since most powerful experimental tool to access electronic
properties of the pseudogap state is angular resolved photoemission
spectroscopy (ARPES)~\cite{ARPES, ARPES1,ARPESYoshida2006}
we performed comparison of LDA+DMFT+$\Sigma_{\bf k}$ calculated
spectral functions and Fermi surfaces with available ARPES 
quasiparticle bands and Fermi surface maps.
Two-particle properties can also be described by this approach~\cite{opt},
e.g. calculated optical spectra in the pseudogap state
compare well with experimental data for Bi2212~\cite{Bi2212} and NCCO~\cite{NCCO_work}.

\section{
LDA+DMFT+$\Sigma$ computational details}
\label{comp}

Crystal structure of Bi2212~\cite{Bi2212}, NCCO~\cite{NCCO_work} and PCCO~\cite{PCCO_work}
has tetragonal symmetry with the space group I4/mmm, while LSCO has ortorhombically distorted
structure Bmab~\cite{LSCO}. For further crystallographic data used within our
LDA+DMFT+$\Sigma_{\bf k}$ approach see Refs.~\cite{Bi2212,LSCO,NCCO_work,PCCO_work}.
Well known quasi two-dimensional nature of these compounds
determines its physical properties.
Physically most interesting are the CuO$_2$ layers. Those layers provide
antibonding Cu-3$d$($x^2\!-\!y^2$) partially filled orbital, whose
dispersion crosses the Fermi level.
Thus we are using this effective LDA calculated Cu-3$d$($x^2\!-\!y^2$) antibonding band as a ``bare''
band in LDA+DMFT+$\Sigma_{\bf k}$ computations. Corresponding hopping integral values
obtained within the linearized
muffin-tin orbitals (LMTO) method~\cite{LMTO} and further application
of the $N$-th order LMTO (NMTO) approach~\cite{NMTO}
are listed in Table~1.

\begin{table}[b]
\label{param}
\caption {Calculated energetic model parameters (eV) and experimental correlation length $\xi$.
First four Cu-Cu in plain hopping integrals
$t$, $t^{\prime}$, $t^{\prime\prime}$, $t^{\prime\prime\prime}$, interplain hopping value~$t_\perp$,
local Coulomb interaction~$U$ and pseudogap potential~$\Delta$.}
\centering
\tiny
\begin{tabular}{|l|c|c|c|c|c|c|c|c|}
\hline
&$t$&$t^{\prime}$&$t^{\prime\prime}$&$t^{\prime\prime\prime}$&$t_\perp$&$U$&$\Delta$&$\xi$ \\
\hline
Bi2212&-0.627& 0.133&0.061& -0.015&0.083&1.51&0.21&10a\\
\hline
NCCO& -0.44&0.153&0.063&-0.01&---&1.1&0.36&50a\\
\hline
PCCO&-0.438&0.156&0.098&---&---&1.1&0.275&50a\\
\hline
LSCO&-0.476&0.077&-0.025&-0.015&---&1.1&0.21&10a\\

\hline
\end{tabular}
\end{table}

Next to perform DMFT calculations one should set up on-site Coulomb interaction values.
The values of Coulomb interaction on effective Cu-3$d$($x^2\!-\!y^2$) orbital $U$ obtained via constrained LDA computations\cite{Gunnarsson} are also presented in the Table~1.

To account for the AFM spin fluctuations, a two-dimensional model 
of the pseudogap state is applied.\cite{Sch,KS}
Corresponding \textbf{k}-dependent self-energy $\Sigma_{\textbf{k}}$~\cite{psgap,Sch,KS}
describes nonlocal correlations induced
by (quasi) static short-range collective
Heisenberg-like AFM spin fluctuations.\cite{lifetime}

The $\Sigma_{\textbf{k}}$ definition contains two important parameters:
the pseudogap energy scale (amplitude) $\Delta$, representing the energy scale
of fluctuating SDW, and the spatial correlation
length $\xi$.
The latter is usually determined from experiment.
The $\Delta$ value was calculated as described in Refs.~\cite{cm05,FNT}.
The value of correlation length was taken in accordance with the typical
value obtained in neutron scattering experiments on NCCO~\cite{NCCOxi} and LSCO~\cite{LSCOxi}.
Employed values of $\Delta$ and $\xi$ for all considered systems are shown in Table~1.
To solve DMFT equations numerical renormalization group (NRG, Refs.~\cite{NRG,BPH}) was
employed as an ``impurity solver''.
Corresponding temperature of DMFT(NRG) computations was 0.011~eV and
hole or electron concentrations were 15\%.

\section{Results and discussion}
\label{results}

\begin{figure}
\includegraphics[clip=true,angle=270,width=0.47\columnwidth]{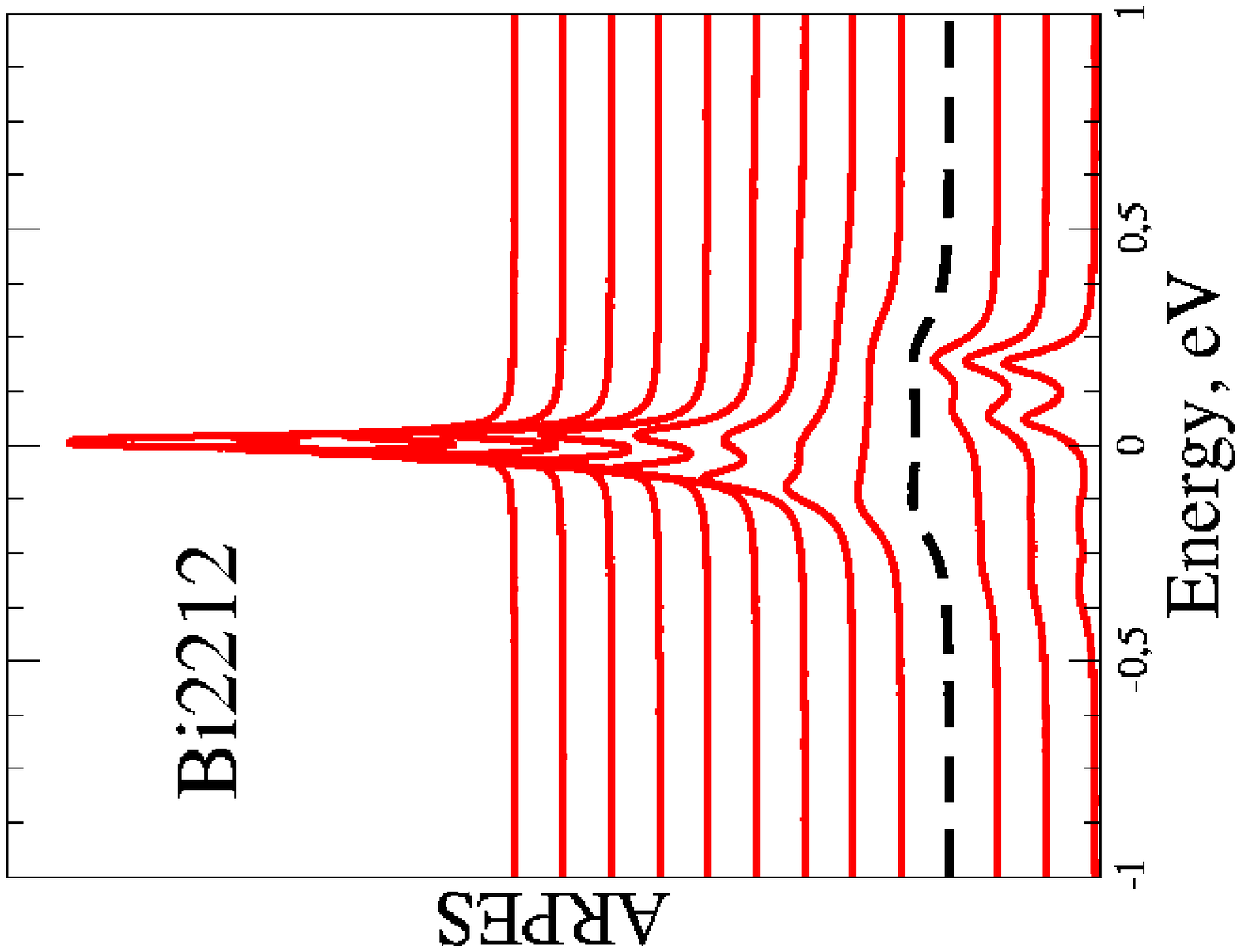}
\includegraphics[clip=true,angle=270,width=0.47\columnwidth]{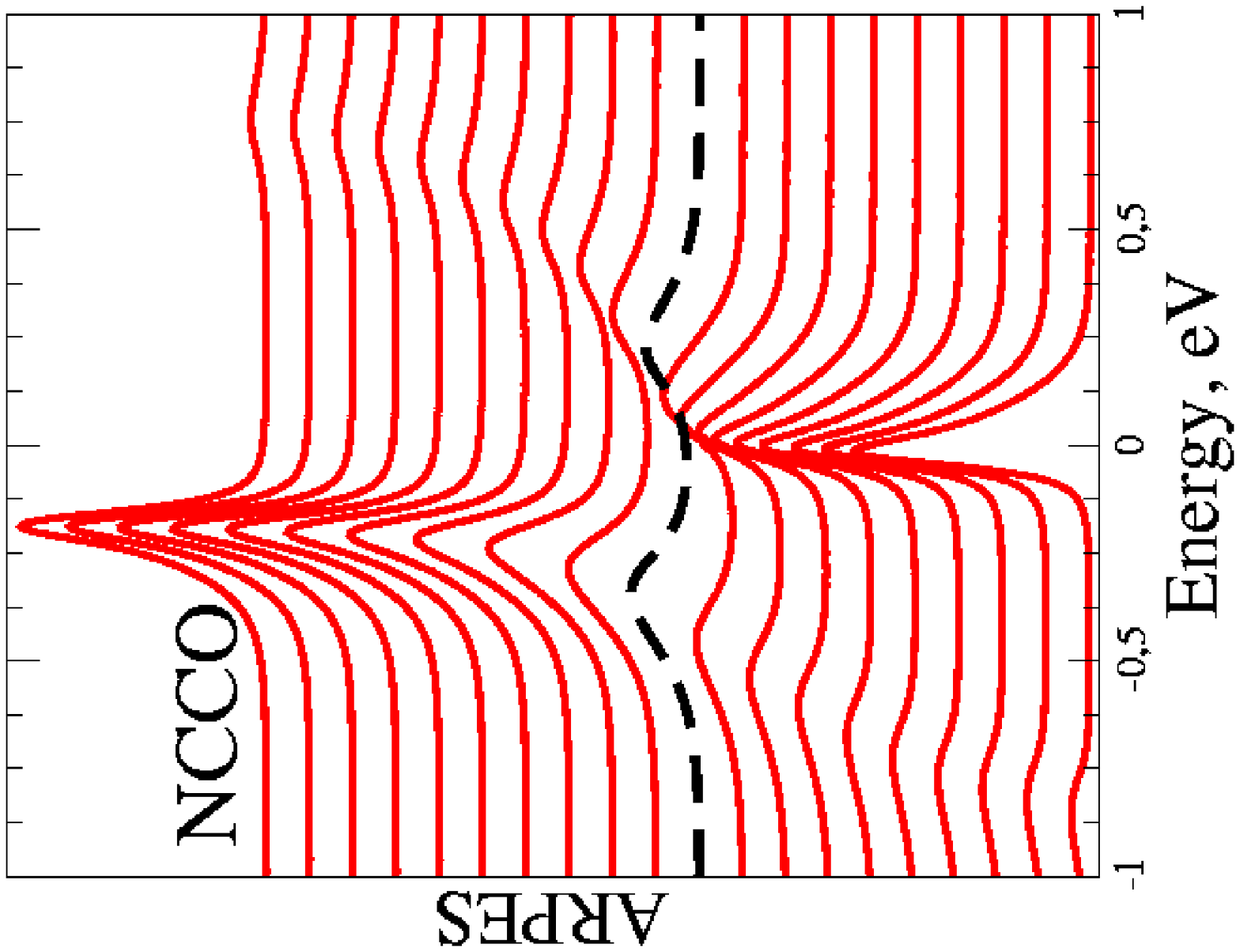}
\caption{LDA+DMFT+$\Sigma_{\bf k}$ spectral functions for Bi2212 (upper panel) and NCCO (lower panel)
along of noninteracting FS in 1/8 of BZ. Dashed-black line corresponds to ``hot-spot'' (Ref.~\cite{NCCO_work})}
\label{sdfs}
\end{figure}

Based on extended analysis of LDA+DMFT+$\Sigma_{\bf k}$ 
results and experimental ARPES data the origin of pronounced ``hot-spots'' 
(cross-point of the Fermi surface and umklapp surface) for electron doped
systems~\cite{NCCO_work,PCCO_work} was established.
Also it was shown that hole doped systems have only Fermi arcs~\cite{Bi2212,LSCO}.
Fig.~\ref{sdfs} displays LDA+DMFT+$\Sigma_{\bf k}$ spectral functions
along 1/8 of noninteracting FS from the nodal point (top curve) to the antinodal
one (bottom curve).
Data for Bi2212 is given in left panel, NCCO --- right panel of Fig.~\ref{sdfs}.
For both compounds  antinodal quasiparticles are well-defined --- sharp peak
close to the Fermi level. Going to the nodal point quasiparticle damping
grows and peak shifts to higher binding energies. This behavior is confirmed
by experiments Refs.~\cite{Armitage02,Kaminski05} (for comparison
with experiment see Ref.~\cite{NCCO_work}). 
Let us interpret the spectral function peaks based on the
LDA+DMFT+$\Sigma_{\bf k}$ results. Namely, for Bi2212 nodal
quasiparticles are formed by low energy edge of pseudogap, while for NCCO they are
formed by higher energy pseudogap edge. Also in NCCO there is obviously
no bilayer splitting effects seen for Bi2212 (left panel of Fig.~\ref{sdfs}).

\begin{figure}[b]
\includegraphics[clip=true,width=1.\columnwidth]{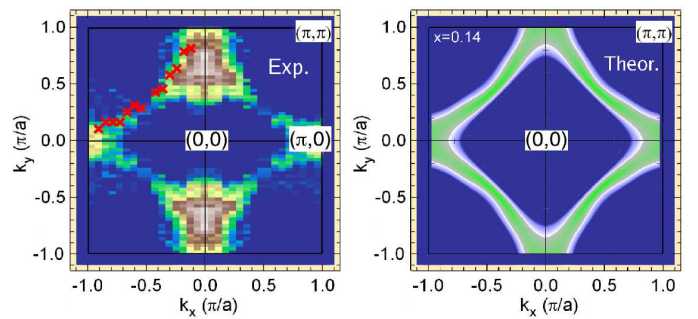}\\
\caption{ Fermi surfaces of LSCO at x=0.14 from experiment (left panel) and LDA+
DMFT+$\Sigma_{\textbf{k}}$
computations (right panel) Red crosses on the left panel correspond to experimental ${\bf k}_F$ values.
(Ref.~\cite{LSCO})
\label{lscofs}}
\end{figure}

``Hot-spots'' for NCCO are
closer to the BZ center~\cite{NCCO_work}. In Fig.~\ref{sdfs} one can see it from the position
of the dashed-black line which corresponds to the ``hot-spot'' ${\bf k}$-point.
For Bi2212 scattering from neighboring BZ amplify each other and instead
of just``hot-spot''  we see rather extended ``destructed'' Fermi surface towards the BZ boundaries.
Such strong scattering comes from scattering processes with momentum transfer of
the order of {\bf Q}=($\pi,\pi$)~\cite{psgap,Sch,KS}, corresponding to AFM pseudogap fluctuations.
Qualitatively the same picture is found also in LSCO (see Fig.~\ref{lscofs}).

Recent experimental and theoretical LDA+DMFT+$\Sigma_{\bf k}$ 
Fermi surface maps~\cite{LSCO} are shown in Fig.~\ref{lscofs}
at panels (a) and (b) correspondingly. Both pictures reveal strong scattering around
($\pi$,0)-point which we associate with scattering in the vicinity of the so-called ``hot-spots''
which are close to the ($\pi$,0)~\cite{Bi2212,NCCO_work}.
Along nodal directions we observe typical Fermi arcs. They are pretty well seen in the theoretical data
while in experiment we obsereve just narrow traces of them
(Bi2212 Fermi surface is compared with experiment in Ref.~\cite{Bi2212}).

\begin{figure}
\includegraphics[clip=true,width=1\columnwidth]{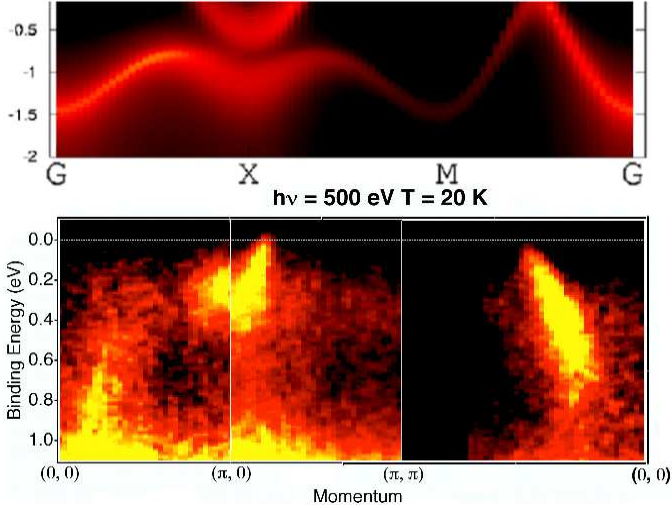}
\caption{Comparison of LDA+DMFT+$\Sigma_{\bf k}$ spectral functions
(upper panel) for NCCO along BZ high-symmetry directions with experimental
ARPES (Ref.~\cite{NCCO_work}) (lower panel).}
\label{ncexp}
\end{figure}

Another possibility to compare LDA+DMFT+$\Sigma_{\bf k}$ results with ARPES data
is spectral function colour maps plotted along symmetry lines.
In Fig.~\ref{ncexp} we present LDA+DMFT+$\Sigma_{\bf k}$ intensity plots along the high symmetry lines
for NCCO  (upper panel) in comparison with
high-energy bulk sensitive angle-resolved photoemission data of Nd$_{1.85}$Ce$_{0.15}$CuO$_4$ (lower
panel).\cite{NCCO_work}
Indeed we see quite a good agreement of LDA+DMFT+$\Sigma_{\bf k}$ and experimental data.
For the $M\!-\!\Gamma$ direction there is not very much going on. Basically we see both
in theory and experiment very intensive quasiparticle band. For the $M\!-\!\Gamma$ direction
less intensive shadow band is not resolved 
in the experiment.

More interesting situation is observed for $\Gamma\!-\!X\!-\!M$ directions.
At $\Gamma$-point there is band in the experiment starting at about -1.2 eV.
It is rather intensive and goes up in energy. Suddenly there is almost zero intensity
at about -0.3 eV. Then in the vicinity of the $X$-point intensity rises up again.
In the $X\!-\!M$ direction around -0.3 eV on the right side of $X$-point there is also quite intensive
region. At a first glance one can think that it is the same band with matrix element effects
governing intensity. However based on analysis of Ref.~\cite{NCCO_work} one can conclude that
this low intensity region is the forbidden gap between
shadow and quasiparticle bands. The ``horseshoe'' around $X$-point is formed by the shadow band on the left and
the quasiparticle band on the right for upper branch and other way round for the lower branch.
As a consequence of that there is also intensive shadow FS sheets around
($\pi/a$,0) point.
Rather intensive nondispersive states at about -1.0 eV within experimental data
can be presumably associated with the lower Hubbard band and
possible admixture of some oxygen states.
Let us also suppose that high intensity at -0.3 eV for $X$ point may be interpreted
not as a van-Hove singularity of bare dispersion but rather of
high-energy pseudogap branch~\cite{NCCO_work}.

\begin{figure}
\includegraphics[clip=true,width=1.\columnwidth]{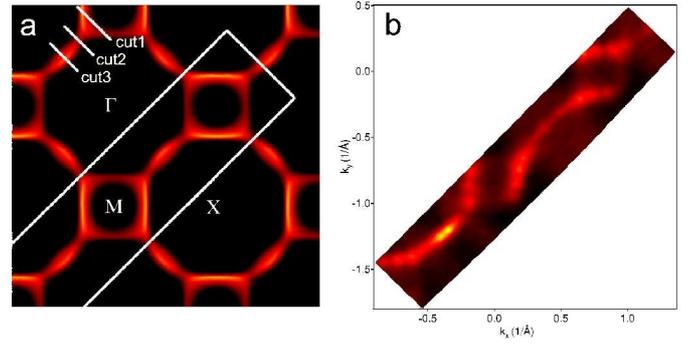}\\
\caption{ (a) Extended Fermi surfaces for PCCO --- LDA+DMFT+$\Sigma_{\textbf{k}}
$ data.
White rectangle on panel (a) schematically shows the part of reciprocal space
measured experimentally (panel b). Lower left corner is X-point ($\pi,0$).(Ref.~\cite{PCCO_work})
\label{FermiSurface}}
\end{figure}

One more fascinating comparison for LDA+DMFT+$\Sigma_{\textbf{k}}$ results with
experimental ARPES data is recently reported by us in the Ref.~\cite{PCCO_work}.
In Fig.~\ref{FermiSurface} an extended picture of PCCO Fermi surfaces is presented
(panel (a) --- LDA+DMFT+$\Sigma_{\textbf{k}}$ results, panel (b) --- experimental ARPES data).
Strictly speaking Fig.~\ref{FermiSurface} is a color map
in reciprocal space of the corresponding
spectral function plotted at  the Fermi level.
FS is clearly visible as reminiscence of non-interacting band close to the first
Brillouin zone border and around
$(\pi/2,\pi/2)$ point (so called Fermi arc), where the quasiparticle
band crosses the Fermi level.
There is an interesting physical effect of  partial ``destruction'' of the FS
observed in the ``hot-spots'', points that are located at  the
intersection of the FS and its AFM umklapp replica.
This FS ``destruction''  occurs because of the  strong electron scattering on the antiferromagnetic
(AFM) spin (pseudogap) fluctuations on the copper atoms.
Also the ``shadow'' FS is visible as it should be for AFM folding.  
As no long-range order is present in the underdoped phase the ``shadow'' FS has
weaker intensity with respect to FS. The PCCO FS is very similar to that of 
Nd$_{2-x}$Ce$_{x}$CuO$_4$ (NCCO), which belongs to the same family of
supperconductors~\cite{NCCO_work,Armitage02}.

\begin{figure}[hb]
\includegraphics[clip=true,width=1\columnwidth]{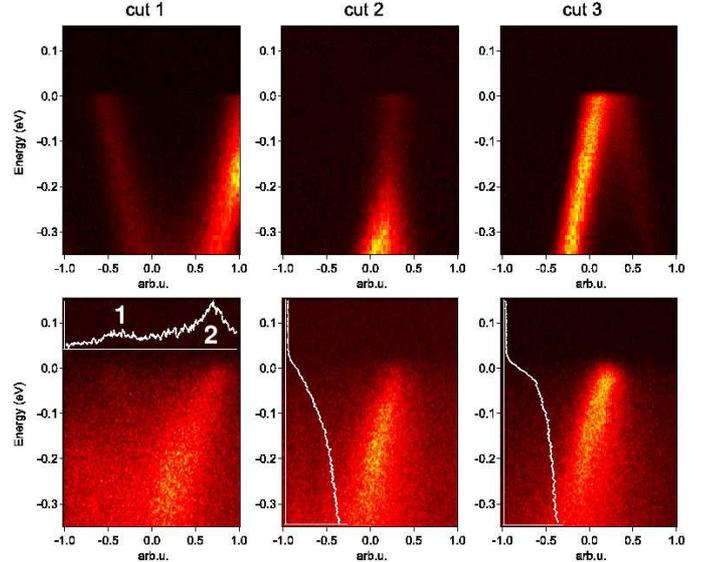}
\caption{Energy--momentum intensity distributions for the specific cuts drawn in
 Fig.~\ref{FermiSurface}
(upper panels --- theoretical data, lower panels --- experimental photoemission
intensity).
To judge about the absolute
intensities of the ``shadow'' (1) and main band (2) cut 1 contains an
MDC curve integrated in an energy window 60 meV
centered at the Fermi level (FL). Similarly integral EDC for cut 2 (``hot-spot'')
shows suppression of the intensity at the
FL as compared to cut 3, which is located further away from
the ``hot-spot''. (Ref.~\cite{NCCO_work})
The FL is zero.\label{Cuts}}
\end{figure}

Let us compare theoretical (upper panels) and experimental (lower panels) energy quasiparticle dispersion
for most characteristic cuts introduced in Fig.~\ref{FermiSurface} (see Fig.~\ref{Cuts}).
Theoretical data were multiplied by the Fermi function at a
temperature of 30K and convoluted
with a Gaussian to simulate the effects of experimental resolution,
with further artificial noise added.

The Cut 1 intersects quasiparticle and ``shadow'' Fermi surfaces close to the Brillouin zone border.
One can find here a ``fork''-like structure formed by the damped ``shadow'' band
 (-0.5-0 arb.u.) and
better defined quasiparticle band (0.5-1 arb.u.). This structure corresponds to
preformation of FS cylinder around ($\pi$,0) point.
The Cut 2 goes exactly through the ``hot-spot''.
Here we see a strong suppression of the quasiparticle band around the
Fermi level similar to NCCO as shown 
in Fig.~\ref{ncexp}. The Cut 3 crosses the Fermi arc, where  we can see a
very well defined
quasiparticle band. However weak intensity ``shadow'' band is also
present. For the
case of long range AFM order and complete folding of electronic structure, FS and its ``shadow''
should form a closed FS sheet
around ($\pi/2$, $\pi/2$) point, while in the current case the part of the pocket
formed by the ``shadow'' band is not
so well defined in momentum space.
As can be seen there is a good correspondence between the calculated
and experimental data in terms of the  above
described behavior, which is also similar
to the results reported for Nd$_{2-x}$Ce$_{x}$CuO$_4$ (NCCO) in our earlier work.\cite{NCCO_work}

\section{Conclusion}
\label{conclusion}

Here we summarize our recent results on LDA+DMFT+$\Sigma_{\textbf{k}}$ investigations of
pseudogap state for a number of copper high-T$_c$ compounds. We considered for the main
prototype systems:  hole doped  -- Bi2212~\cite{Bi2212} and LSCO~\cite{LSCO};
electron doped -- PCCO~\cite{PCCO_work} and NCCO~\cite{NCCO_work}.
For all compounds the LDA+DMFT+$\Sigma_{\textbf{k}}$ calculations show that Fermi-liquid behavior is still
conserved far away from the ``hot-spots'' (antinodal direction),
while the  destruction of the Fermi surface observed in the vicinity of
``hot-spots'' (close to nodal direction). This destruction is due to strong scattering of correlated electrons
on short-range antiferromagnetic (pseudogap) fluctuations.
Moreover the origin of clearly observed ``hot-spots'' for electron doped systems 
(in contrast to hole doped ones with a Fermi arcs only) is established.
Comparison between experimental ARPES and
LDA+DMFT+$\Sigma_{\textbf{k}}$ data  reveals a
good semiquantitative agreement.
The experimental and theoretical results obtained once again support the
AFM scenario of pseudogap formation not only in hole doped HTSC systems\cite{Bi2212,LSCO} but also
in electron doped ones~\cite{NCCO_work}.

\section{Acknowledgments} 


This work is partly supported by RFBR grant 08-02-00021 and was performed
within the framework of programs of fundamental research of the Russian Academy
of Sciences (RAS) ``Quantum physics of condensed matter'' (09-$\Pi$-2-1009) and
of the Physics Division of RAS  ``Strongly correlated electrons in solid states''
(09-T-2-1011). IN thanks Grant of President of Russia MK-614.2009.2,
interdisciplinary UB-SB RAS project, and Russian Science Support
Foundation.



\begin{thebibliography}{99}
\bibitem{lifetime}The quasi static approximation for AFM fluctuations necessarily limits
our approach to high - enough temperatures (energies not so close to the Fermi
level)\cite{Sch,KS}, so that in fact we are unable to judge e.g. on 
the nature of low temperature (energy) damping in our model.
\end{thebibliography}
\end {document}